\documentstyle[preprint,aps]{revtex}
\tightenlines
\begin{document}
\draft
\title{\bf Semiflexible polymer on an anisotropic Bethe lattice}
\author{J.F. Stilck\cite{l}, C.E. Cordeiro, and R.L.P.G. do Amaral}
\address{
Instituto de F\'\i sica,\\
Universidade Federal Fluminense\\
Av. Litor\^anea s/n\\
24210-340, Niter\'oi, R.J., Brazil\\}
\date{\today }
\maketitle
\begin{abstract}

The mean square end-to-end distance of a $N$-step polymer on a Bethe
lattice is calculated. We consider semiflexible polymers placed on
isotropic and anisotropic lattices. The distance on the Cayley tree
is defined by embedding the tree on a sufficiently high dimensional
Euclidean space considering that every bend of the polymer defines a
direction orthogonal to all the previous ones. In the isotropic case,
the result obtained for the mean square end-to-end distance turns out
to be identical to the one obtained for {\em ideal} chains without
immediate returns on an hypercubic lattice with the same coordination
number of the Bethe lattice. For the general case, we obtain the
asymptotic behavior in the semiflexible and also in the almost rigid
limits.
\end{abstract}
\pacs{05.50.+q; 61.41+e}

\section{INTRODUCTION}
\label{s1}
Chain polymers are often approximated as self- and mutually avoiding
walks (SAW's) on a lattice, and much information about the behavior
of polymers both in a melt or in solution has been understood
theoretically through this model \cite{f53,dg79}. One of the
characterizations of the conformations of a walk is through its
mean-square end-to-end distance $\langle R^{2}\rangle $, where the mean is taken
over all configurations of the $N$-step walk on the lattice. In the
limit $N\rightarrow \infty$ a scaling behavior $\langle R^{2}\rangle \sim
N^{2\nu}$ is observed, where the exponent $\nu$ exhibits universal
behavior, with a mean-field value $\nu=1/2$ for random walks, which
correspond to ideal chains, and $\nu=3/4$ for SAW's on two-dimensional
lattices \cite{n82}, for example.

An interesting question arises if the chains are not considered to be
totally flexible, an energy being associated to bends of the chain.
This is often observed for real polymers. Let us, for simplicity,
restrict ourselves to SAW'S on hypercubic lattices. In this case,
consecutive steps of the walk are either in the same direction or
perpendicular. So, a Boltzmann factor $z$ may be associated to each
pair of perpendicular consecutive steps of walk. This problem of
semiflexible polymers (also called persistent or biased walks) has
been studied for some time\cite{ts75,mn91,cfs92}, and there occurs a
crossover in the behavior of the walk between a rodlike behavior
$\nu_{r}=1$ for $z=0$, where the polymer is totally stiff, and the
usual behavior with a different exponent $\nu$ for nonzero values of
$z$. Stating this point more precisely, the mean square end-to-end
distance displays scaling behavior in the limit  $N\rightarrow
\infty;z\rightarrow0;N^{\psi}z=constant$, which is given by
\begin{equation}
\langle R^{2}\rangle \sim N^{2\nu_{r}}F(zN^{\psi}),
\label{e1}
\end{equation}
the observed values being $\nu_{r}=1$ and $\psi=1$.

The scaling function has the behavior $F(x)\sim x^{(2\nu-2)/\psi}$
in the limit $x \rightarrow \infty$.  This scaling form has been
verified through several techniques, although in three dimensions a
mean field exponent $\nu=1/2$ was found for intermediate values of
the number of steps $N$, the crossover to the three-dimensional value
occurring at rather high values of $N$\cite{mn91}.

In this paper, we consider the problem of a semiflexible polymer on a
Bethe lattice\cite{b82}, calculating exactly the mean square
end-to-end distance of walks on the Cayley tree which start at the
central site and have $N$ steps, supposing that the walks will never
reach the surface of the Cayley tree, thus remaining in its core. We
calculate also the mean square end-to-end distance in the case when
the lattice is considered  anisotropic, that is, when the edges of
the lattice are not equivalent with respect to their occupation by a
polymer bond. The definition of the distance between two sites of the
Cayley tree is not obvious, and some possibilities exploring the fact
that the tree may be embedded in a hypersurface of a non-Euclidean
space have been given\cite{m92}. In this paper, however, we used a
simpler definition, considering the Cayley tree in the thermodynamic
limit to be embedded in an infinite-dimensional Euclidean space. The
result for $\langle R^{2}\rangle (N,z)$, for the isotropic case has
the scaling form of Eq. \ref{e1}. Not surprisingly the scaling
function $F(x)$ is equal to the one obtained for random walks with no
immediate return on an hypercubic lattice with the same coordination
number of the Bethe lattice considered.  This might be expected since
Bethe lattice calculations lead to mean-field critical exponents.
Also, in the limit $N\rightarrow\infty$ for nonzero values of $z$ the
scaling behavior $\langle R^{2}\rangle \sim N^{2\nu}$ with the
classical value $\nu=1/2$ is verified in the expression for $\langle
R^{2}\rangle (N,z)$. It should be mentioned that our proposal of
defining the Euclidean distance between two points of the Cayley tree
is similar to earlier results in the literature relating this
distance to the chemical distance, measured along the chain
\cite{dq86}. However, the distinction between the chemical and 
the Euclidean distances is not always properly considered 
in the literature, and this may lead to
contradicting results \cite{hi98}, as we will discuss in more detail
in the conclusion.

In section \ref{s2} we define the model and calculate the mean square
end-to-end distance recursively on the anisotropic Bethe lattice. The
problem is then reduced to finding the general term of a {\em linear}
mapping in six dimensions. In the particular case of an
isotropic lattice, we find a closed expression for
$\langle R^2\rangle $. In section \ref{s3} the asymptotic behavior is
studied for the general case, based on the mapping. In section
\ref{s4} final comments and discussions may be found. Finally, we
present in the appendices a combinatorial calculation for $\langle
R^2\rangle$ in the isotropic case.

\section{DEFINITION OF THE MODEL AND SOLUTION FOR THE ISOTROPIC
LATTICE}
\label{s2}
We consider a Cayley tree of coordination number $q$ and place a
chain on the tree starting at the central site. Each bond of the
tree is supposed to be of unit length. Figure \ref{f1} shows
a tree with $q=4$ and a polymer with $N=2$ steps placed on it. Since
we want the Cayley tree to be an approximation of a hypercubic
lattice in $d$ dimensions, we will restrict ourselves to even
coordination numbers $q=2d$. As in the hypercubic lattice, the bonds
incident on any site of the tree are in $d$ directions, orthogonal to
each other. As may be seen in Figure \ref{f1}, the central site of
the tree is connected to $q$ other sites, which belong to the first
generation of sites. Each of the sites of the first generation is
connected to $(q-1)$ sites of the second generation, and this process
continues until the surface of the tree is reached, after a number of
steps equal to the number of generations in the tree. Upon reaching a
site of the $i$'th generation coming from a site belonging to
generation $(i-1)$, there are $(q-1)$ possibilities for the next step
of the walk towards a site of generation $(i+1)$. One of them will be
in the same direction as the previous step, while the remaining
$(q-2)$ will be in directions orthogonal to {\em all} previous steps.
In the second case, a statistical weight $z$ is associated to the
elementary bend in the walk. Therefore, we admit that the $(q-2)$
bonds which are orthogonal to the last step are also orthogonal to
{\em all} bonds of the lattice in earlier generations. Let us stress
two consequences of this supposition: (i) A tree of coordination
number $q$ with $N_{g}$ generations will be embedded in a space of
dimension
\begin{equation}
D=q/2+(N_{g}-1)(q/2-1).
\label{e3}
\end{equation}
The sites of the Cayley tree will all be sites of a hypercubical
lattice in $D$ dimensions. This may be seen in Figure \ref{f1} where
the sites of a tree with $q=4$ and $ N_{g}=2$ are sites of a cubic
$(D=3)$ lattice. (ii) By construction, there will never be loops in
the tree, a property which is true for any Cayley tree. It is well
known
\cite{b82} that it may be shown by other means that the Cayley tree
is a infinite-dimensional lattice in the thermodynamic limit
$N_{g}\rightarrow\infty$. Finally, the anisotropy is introduced into
the model considering that {\em bonds} of the chain in $s$ of the $q/2$
directions (we will call them special) at each lattice site 
contribute with a factor $y$ to the partition function, while no
additional contribution comes from bonds in the remaining $t=q/2-s$
directions at each lattice site.

Usually \cite{b82}, the calculation of thermodynamic properties of
models defined on the Bethe lattice is done in a recursive manner, so
we will follow a similar procedure in the calculation of the mean
square end-to-end distance. We define a generalized partition (or
generating) function for $N$-step chains
\begin{equation}
g_N=\sum z^m y^{N_e} p^{R^2},
\label{e4}
\end{equation}
where the sum is over all configurations of the chain, $z$ is the
statistical weight of an elementary bend in the chain, $y$ is the
statistical weight of bonds in special directions and $p$ is a
parameter associated to the square of the end-to-end distance of the
chain. At the end of the calculation, we will take $p=1$. The numbers
of elementary bends, bonds in special directions, and the square
end-to-end distance of each chain are $m$, $N_e$, and $R^2$,
respectively. The mean
square end-to-end distance may then be calculated through
\begin{equation}
\langle R^2 \rangle _N=\frac{1}{g_N}\left.\left(p\frac{\partial
g_N}{\partial p}\right)\right|_{p=1}.
\label{e5}
\end{equation}

The partition function may then be calculated in a recursive way if
we define partial partition functions $a_N^l$ and $b_N^l$ such that
the first ones include all $N$-bond chains whose last $l$ bonds are
collinear and in one of the special directions (there is necessarily a
bend before the $l$ bonds, if $l<N$), while the last $l$ bonds of the
chains contributing to $b_N^l$ are collinear and in one of the
non-special directions. The partition function may then be written as
\begin{equation}
g_N=\sum_{l=1}^{N} (a_N^l+b_N^l).
\label{e6}
\end{equation}
Due to the fact that there are no closed loops on the Cayley tree, it
is quite easy to write down recursion relations for the partial
partition functions
\begin{mathletters}
\label{e7}
\begin{eqnarray}
a_{N+1}^1&=&2syzp \sum_{l=1}^N b_N^l + 2(s-1)yzp \sum_{l=1}^N
a_N^l,\\ 
a_{N+1}^{l+1}&=&yp^{2l+1}a_N^l,\\
b_{N+1}^1&=&2tzp\sum_{l=1}^N a_N^l + 2(t-1)zp\sum_{l=1}^N b_N^l,\\
b_{N+1}^l&=&p^{2l+1}b_N^l,
\end{eqnarray}
\end{mathletters}
with the initial conditions
\begin{mathletters}
\label{e8}
\begin{eqnarray}
a_1^1&=&2syp,\\
b_1^1&=&2tp.
\end{eqnarray}
\end{mathletters}
For example, in the first expression above, the new bond may be
preceded by a bond in a special direction, with $2s$
possibilities, or by a bond in a non-special direction, with $2(s-1)$
possibilities. In both cases, a factor $p$ is present since the bond
added is in a direction perpendicular to all previous ones, and thus
$R^2$ is increased by one unit. Finally, the inclusion of the new bond
introduces one bend in the chain, thus explaining the factor $z$, and
since the bond is in a special direction the factor $y$ is justified.
In the second expression, it should be mentioned that $R^2$ is
increased by $(l+1)^2-l^2$, thus explaining the exponent of $p$. If
we now define
\begin{mathletters}
\label{e9}
\begin{eqnarray}
a_N=\sum_{l=1}^N a_N^l,\\
b_N=\sum_{l=1}^N b_N^l,
\end{eqnarray}
\end{mathletters}
the mean square end-to-end distance will be
\begin{equation}
\langle R^2 \rangle_N=\frac{1}{a_N+b_N}\left.\left[ p\frac{\partial}
{\partial p}(a_N+b_N) \right] \right|_{p=1} = \frac{c_N+d_N}
{a_N+b_N}. 
\label{e10}
\end{equation}
The recursion relations for $a_N$ and $b_N$, as well as the ones for
the new variables $c_N$ and $d_N$ may be written, for $p=1$, as
\begin{mathletters}
\label{e11}
\begin{eqnarray}
a_{N+1}&=&2z\sum_{l=0}^N y^{l+1}\left[ (s-1)a_{N-l}+sb_{N-l} \right],\\
b_{N+1}&=&2z\sum_{l=0}^N \left[ ta_{N-l}+sb_{N-l} \right],\\
c_{N+1}&=&2z\sum_{l=0}^N y^{l+1} \left[ (s-1)(l+1)^2 a_{N-l}+s(l+1)^2
b_{N-l} +(s-1)c_{N-l}+sd_{N-l} \right],\\
d_{N+1}&=&2z \sum_{l=0}^N \left[ t(l+1)^2 a_{N-l}+(t-1)(l+1)^2 b_{N-l}
+ tc_{N-l}+(t-1)d_{N-l} \right],
\end{eqnarray}
\end{mathletters}
with the initial conditions
\begin{mathletters}
\label{e12}
\begin{eqnarray}
a_0&=&\frac{2s}{z(q-2)},\\
b_0&=&\frac{2t}{z(q-2)},\\
c_0&=&d_0=0
\end{eqnarray}
\end{mathletters}
An undesirable feature of the recursion relations Eqs. \ref{e11} is that
the new values of the iterating variables depend on all previous
values. This dependence, however, is rather simple, and it is
possible, introducing two more variables $e_N$ and $f_N$, to rewrite
the recursion relations as a mapping involving only one previous
value of each variable, valid for $N \ge 1$
\begin{mathletters}
\label{e13}
\begin{eqnarray}
a_{N+1}&=&ya_N+2zy\left[ (s-1)a_N+sb_N \right],\\
b_{N+1}&=&b_N+2z \left[ ta_N+(t-1)b_N \right],\\
c_{N+1}&=&yc_N+2zy\left[ (s-1)(a_N+c_N)+s(b_N+d_N) \right] +(2N+1)ya_N
- 2ye_N,\\
d_{N+1}&=&d_N+2z \left[ t(a_N+c_N)+(t-1)(b_N+d_N) \right] + (2N+1)b_N -
2f_N,\\
e_{N+1}&=&ye_N+2Nzy \left[ (s-1)a_N+sb_N \right],\\
f_{N+1}&=&f_N+2Nz \left[ ta_N+(t-1)b_N \right],
\end{eqnarray}
\end{mathletters}
with the initial conditions
\begin{mathletters}
\label{e14}
\begin{eqnarray}
a_1&=&c_1=2sy,\\
b_1&=&d_1=2t,\\
e_1&=&f_1=0.
\end{eqnarray}
\end{mathletters}
The value for $\langle R^2 \rangle$ may be found iterating the
mapping above through the Eq. \ref{e10}. In principle, since the
mapping is {\em linear}, it is solvable. One starts finding the
general term of the first two equations, then solving the last two
and finally solving the two remaining relations. A software for
algebraic computing is helpful, but we realized that the general
answer will be too large to be handled, and also the computer time
and memory required are beyond the resources we have available. We
therefore restrict ourselves to a complete solution of the isotropic
case $y=1$ and to an exact study of the asymptotic properties of the
solution for the general case. It is worthwhile to observe in the
mapping Eqs. \ref{e13} that under a transformation
\begin{mathletters}
\label{e15}
\begin{eqnarray}
s^{\prime}=t,\\
t^{\prime}=s,\\
y^{\prime}=1/y,
\end{eqnarray}
\end{mathletters}
$\langle R^2 \rangle$ will be invariant, as expected.

For the isotropic case ($y=1$) the mapping Eqs. \ref{e13} is reduced to
three variables
\begin{mathletters}
\label{e16}
\begin{eqnarray}
\alpha_N&=&a_N+b_N,\\
\beta_N&=&c_N+d_N,\\
\gamma_N&=&e_N+f_N,
\end{eqnarray}
\end{mathletters}
and it may be written as
\begin{mathletters}
\label{e17}
\begin{eqnarray}
\alpha_{N+1}&=&\left[ 1+z(q-2) \right] \alpha_N,\\
\beta_{N+1}&=&\left[ 1+z(q-2)\right] \beta_N+ \left[ 2N+1+z(q-2)
\right] \alpha_N - 2\gamma_N,\\
\gamma_{N+1}&=& \gamma_{N}+Nz(q-2)\alpha_N.
\end{eqnarray}
\end{mathletters}
The initial conditions are
\begin{eqnarray}
\alpha_1&=&\beta_1=q,\\
\gamma_1&=&0.
\label{e18}
\end{eqnarray}
It is easy to find the general solution for this mapping
\begin{mathletters}
\label{e19}
\begin{eqnarray}
\alpha_N&=&qk^{N-1},\\
\beta_N&=&\frac{q}{(k-1)^2} \left[ N(k^2-1)k^{N-1}+2-2k^N \right],\\
\gamma_N&=&\frac{q}{k-1} \left[ N(k^2-1)k^{N-1}+1-k^N \right],
\end{eqnarray}
\end{mathletters}
where $k=1+(q-2)z$. The substitution of this solution in Eq. \ref{e10}
results in
\begin{equation}
\langle R^{2}\rangle
=\frac{2[1+a]}{a^{2}}\left[Na-1+\frac{1}{[1+a]^{N}}\right]-N,
\label{e20}
\end{equation}
where $a=k-1=(q-2)z$.

The properties of the mean square end-to-end distance Eq. \ref{e20}
in some limiting cases show that our result has the expected
behavior. First, we notice that when the bend statistical weight $z$
vanishes we have
\begin{equation}
lim_{z\rightarrow0}\langle R^{2}\rangle =N^{2},
\label{e21}
\end{equation}
for any number of steps $N$. This rodlike behavior is expected, since
no bend will be present in the walk.  In the opposite limit of
infinite bending statistical weight the result is
\begin{equation}
lim_{z\rightarrow\infty}\langle R^{2}\rangle =N,
\label{e22}
\end{equation}
which is also an expected result, since in this limit there is a bend
at every internal site of the chain, so that, according to the
definition of the end-to-end distance we are using, the vector
$\vec{R}$ in this situation will have $N$ components, all of them
being equal to 1.

In the limit of an infinite chain $N\rightarrow\infty$ we get, for
nonzero $z$,
\begin{equation}
lim_{N\rightarrow\infty}\langle R^{2}\rangle =\frac{(2+a)N}{a},
\label{e23}
\end{equation}
and we notice that the expected scaling behavior $\langle
R^{2}\rangle \sim N^{2\nu}$ is obtained with the mean field
exponent $\nu=1/2$. The asymptotic behavior of $\langle R^{2}\rangle
$ is different for zero and nonzero $a$, as may be appreciated
comparing Eqs. \ref{e21} and
\ref{e23} respectively. So we may look for the crossover between both
behaviors in the limit of Eq. \ref{e1}, getting the result
\begin{equation}
lim_{N\rightarrow\infty;a\rightarrow0;aN=x}\langle R^{2}\rangle =N^{2}F(x),
\label{e23p}
\end{equation}
with a scaling function
\begin{equation}
F(x)=\frac{2(x-1+\exp(-x))}{x^{2}}.
\label{e24}
\end{equation}
It should be stressed that the square end-to-end distance given in
Eq. \ref{e20} is the same obtained by adapting the general result of
Flory for random walks without immediate return \cite{f53} to
hypercubic lattices. In general, it may be shown that the exact
solution of statistical models with first neighbor interactions on
the Bethe lattice is equivalent to the Bethe approximation on the
Bravais lattice with the same coordination number \cite{b82}. The
random walk without immediate returns corresponds to the Bethe
approximation of the $n \rightarrow 0$ model associated to the
self-avoiding walk problem \cite{sw87}, and here we show that the
analogy may be extended to the mean square end-to-end distance if we
define distances on the Bethe lattice as was done above. Although the
results on the Bethe lattice as calculated here and the ones for
ideal chains without immediate return on a hypercubic lattice with
the same coordination number should have the same asymptotic
behaviors, it is at first surprising that they are actually
identical. However, it turns out that the mean value of the angle
between successive bonds, as calculated by Flory in his original work
\cite{f53}, is actually {\em exact} for chains on the Bethe lattice
as we considered.

\section{ASYMPTOTIC BEHAVIOR IN THE GENERAL CASE}
\label{s3}
In this section we develop a study of the asymptotic solution of the
mapping Eqs. \ref{e13} for $N \gg 1$. Let us
reduce the dimension of the mapping by one defining new iteration
variables
\begin{mathletters}
\label{e25}
\begin{eqnarray}
B_N=\frac{b_N}{a_N},\\
C_N=\frac{c_N}{a_N},\\
D_N=\frac{d_N}{a_N},\\
E_N=\frac{e_N}{a_N},\\
F_N=\frac{f_N}{a_N}.
\end{eqnarray}
\end{mathletters}
From Eqs. \ref{e13} and the initial conditions Eqs. \ref{e14} it is easy
to write the recursion relations and initial conditions for the new
iterative variables in the mapping above. In the limit of large
values of $N$, for fixed $z$ and $y$,  the following asymptotic
behavior is observed
\begin{mathletters}
\label{e26}
\begin{eqnarray}
B_N &\sim& B^0,\\
C_N &\sim& C^0+C^1N,\\
D_N &\sim& D^0+D^1N,\\
E_N &\sim& E^0+E^1N,\\
F_N &\sim& F^0+F^1N.
\end{eqnarray}
\end{mathletters}
The substitution of the these expressions into the recursion
relations for the variables defined in Eqs. \ref{e25}, obtained from the general
mapping Eqs. \ref{e13}, leads to the determination of the constants in the
asymptotic behavior and thus we obtain
\begin{equation}
\langle R^2 \rangle =\frac{C_N+D_N}{1+B_N} \sim \frac{C^1+D^1}
{1+B^0} N = C N,
\label{e27}
\end{equation}
where the amplitude $C=C^1$ is given by
\begin{equation}
C=\frac{sy(B^0)^2 \left[ \frac{y(1+\epsilon)+1}{y(1+\epsilon)-1}
\right] +t\left[ \frac{2+\epsilon}{\epsilon}\right] }{sy(B^0)^2+t},
\label{e28}
\end{equation}
where
\begin{equation}
\epsilon=2z(s-1+sB^0),
\label{e29}
\end{equation}
and $B^0$ is the positive root of
\begin{equation}
2zsy(B^0)^2+\left[ y-1+2z(sy-t-y+1)\right] B^0-2zt=0.
\label{e30}
\end{equation}
The amplitude of the asymptotic behavior of $\langle R^2 \rangle$ thus
may be obtained exactly in the general case and, as may be seen in
Figure \ref{f2}, diverges as $z \rightarrow 0$, as expected. Also,
in the limit $y \rightarrow \infty$ the problem reduces to a walk on
an isotropic lattice with coordination number equal to $2s$, and
we get
\begin{equation}
C=\frac{2z(s-1)+2}{2z(s-1)},
\label{e31}
\end{equation}
which agrees with Eq. \ref{e19} for the isotropic case.

Now we will study the asymptotic behavior in the quasi-rigid limit $N
\rightarrow \infty$, $z \rightarrow 0$, and $N(q-2)z=x$. We thus expand
$\langle R^2 \rangle$ for small values of $z$
\begin{equation}
\langle R^2 \rangle (z,y,N) \sim \langle R^2 \rangle (0,y,N) +
\left.\frac{\partial \langle R^2 \rangle}{\partial z}\right|_{z=0} z.
\label{e32}
\end{equation}
For $z=0$ the solution of the mapping Eqs. \ref{e13} is
\begin{mathletters}
\label{e33}
\begin{eqnarray}
a_N&=&2sy^N,\\
b_N&=&2t,\\
c_N&=&2sy^NN^2,\\
d_N&=&2tN^2,\\
e_N&=&f_N=0;
\end{eqnarray}
\end{mathletters}
and we have $\langle R^2 \rangle=N^2$, as expected. From the mapping
Eqs. \ref{e13} the recursion relations for the derivatives of
the variables with respect to $z$ (at $z=0$) may be seen to be
\begin{mathletters}
\label{e34}
\begin{eqnarray}
a^\prime_{N+1}&=&ya^\prime_N+4ys\left[(s-1)y^N+t\right],\\
b^\prime_{N+1}&=&b^\prime_N+4t(sy^N+t-1),\\
c^\prime_{N+1}&=&yc^\prime_N+4ys(1+N^2)\left[(s-1)y^N+t\right] +y(2N+1)
a^\prime_N-2ye^\prime_N,\\
d^\prime_{N+1}&=&d^\prime_N+4t(1+N^2)(sy^N+t-1)+(2N+1)b^\prime_N -
2f^\prime_N,\\ 
e^\prime_{N+1}&=&ye^\prime_N+4Nys\left[(s-1)y^N+t \right],\\
f^\prime_{N+1}&=&f^\prime_N+4Nt(sy^N+t-1),
\end{eqnarray}
\end{mathletters}
where the values for the variables (Eqs. \ref{e33}) have already been
substituted and the initial conditions are
$a^\prime_1=b^\prime_1=...= f^\prime_1=0$. The general solution of
the recursion relations Eqs. \ref{e34} is not difficult to obtain with the
aid of an algebra software. Considering the invariance described in Eqs. \ref{e15} we
will restrict our discussion to the case $y>1$, without loss of
generality. For large values of $N$, the dominant terms of the
solution of the mapping are
\begin{mathletters}
\label{e35}
\begin{eqnarray}
a_N+b_N &\sim& 2sy^N,\\
c_N+d_N &\sim& 2sy^N N^2,\\
a^{\prime}_N+b^{\prime}_N &\sim& \left\{
\begin{array}{ll}
4s(s-1)y^N N & \mbox{if $s>1$}\\
\frac{8ty^N}{y-1} & \mbox{if $s=1$}
\end{array}
\right. ,\\
c^{\prime}_N+d^{\prime}_N &\sim& \left\{
\begin{array}{ll}
\frac{8}{3}s(s-1)y^N N^3 & \mbox{if $s>1$}\\
\frac{8ty^N N^2}{y-1} & \mbox{if $s=1$}
\end{array}
\right. .
\end{eqnarray}
\end{mathletters}
The leading term in the derivative of the mean-square end-to-end
distance will be
\begin{equation}
\left. \frac{\partial \langle R^2 \rangle}{\partial z}\right|_{z=0}
\sim -\frac{2}{3}(s-1)N^3.
\label{e36}
\end{equation}
Therefore, up to first order in $x$, the scaling function $F(x)$ in
the quasi-rigid limit is found to be $F(x) \sim 1 - F_1(s,t)x$. 
Considering the symmetry Eq. \ref{e15} and
the solution for the isotropic case Eq. \ref{e20}, we have
\begin{equation}
F_1(s,t)=\left\{
\begin{array}{lll}
\frac{2(t-1)}{3(q-2)} & \mbox{if $y<1$},\\
\frac{1}{3} & \mbox{if $y=1$},\\
\frac{2(s-1)}{3(q-2)} & \mbox{if $y>1$}.
\end{array}
\right. 
\label{e37}
\end{equation}
We thus conclude that the scaling function in general displays a
discontinuous derivative at $y=1$.

\section{CONCLUSION}
\label{s4}
We formulated the problem of the calculation of the mean square
end-to-end distance of semiflexible polymers placed on a
$q$-coordinated anisotropic Bethe lattice as a linear mapping, whose
general term may in principle be obtained. In the isotropic case, the
mapping may easily be solved and leads to an expression for $\langle
R^2 \rangle$ which is {\em identical} to the one obtained for random
walks without immediate return on a hypercubic lattice with the same
coordination number \cite{f53}. The identity between the two problems
regarding thermodynamic properties derived from the free energy is
well known \cite{sw87}, and is here extended for a thermodynamic
average of a geometric property. One point which should be stressed
is that the definition of the Euclidean distance between two points
on the Bethe lattice is rather arbitrary. Here we defined the
distance by embedding the Cayley tree in a hypercubic lattice of
sufficiently high dimensionality. In the thermodynamic limit the
dimensionality of this lattice diverges, as expected \cite{b82}.
Other definitions of distance may be proposed \cite{m92}. The simple
one we adopted here leads to meaningful conclusions. Since
calculations on the Bethe lattice are usually done recursively, and
one step in the recursion relations corresponds to adding another
generation to the tree, it is tempting to define the distance between
two sites on the tree as the difference between the numbers of the
generations they belong to. This definition, although simple and
operational, has serious drawbacks, however. This is quite clear for
the particular problem we looked at here, since it implies that
$\langle R^2 \rangle$ for {\em any} $N$-step chain is equal to $N^2$.
We would thus have $\nu=1$, the one-dimensional value, and the
identity between the results for the Bethe lattice and for walks
without immediate return on hypercubic lattices would break down.
This definition of distance was used recently in the exact
calculation of correlation functions for a general spin-$S$ magnetic
model \cite{hi98}, leading to $\nu=1$, in opposition to the generally
accepted mean-field value $\nu=1/2$ \cite{tm96}.

The fact that all walks we considered here have their initial site
located at the central site of the Cayley tree is of course
convenient for the calculations and may be seen as a particular case.
A closer consideration of this point, however, leads to the
conclusion that our results are exact for any chains such that the
assertion that at any bend the new direction is perpendicular to {\em
all} previous directions of bonds holds. Thus, it is clear that if
the whole chain is contained in one of the $q$ rooted sub-trees
attached to the central site the results are still the same. If
portions of the chain are located on two of these sub-trees the
calculation becomes more complicated since, as may be seen in figure
\ref{f1}, there are bonds in the same direction in different
sub-trees. However, this problem may be easily avoided by enlarging
the dimension of the euclidean space in which the tree is embedded,
thus assuring that any two bonds in the same direction are
necessarily connected by a walk without any bend. For such a tree,
our results hold for any chain, regardless of the location of its
endpoints. 

In the general anisotropic case, we restricted ourselves to the
discussion of the asymptotic behavior of $\langle R^2 \rangle$, which
was studied in the semiflexible case and also in the quasi-rigid
limit. The expected scaling behavior was obtained in both cases, and
a interesting discontinuity in the quasi-rigid limit amplitude is
observed as the isotropic value $y=1$ is crossed.

\acknowledgments
We acknowledge partial financial support  from the
Brazilian agencies CNPq and FINEP.

\appendix
\section{COMBINATORIAL SOLUTION IN THE ISOTROPIC CASE}
\label{a1}
 
Any $N$-step walk on the Cayley tree will visit a subset of sites of
the D-dimensional hypercubic lattice defining a subspace whose
dimensionality is between 1 and $N$. The limiting cases are the ones
of a polymer without any bend (rod), which is one-dimensional, and a
polymer where we have a bend at every internal site, and since at
each bend the new bond is in a direction orthogonal to all precedent
bonds of the polymer, the polymer is embedded in a $N$-dimensional
subspace. Since the initial site of the chain is supposed to be at
the central site of the tree, the end-to-end distance will be given
by the modulus of the position vector of the final site, denoted by
$\vec{R}$. For a polymer with $m$ bends, the number of components of
this vector will be equal to $m+1$. For simplicity, we will admit
that each bond is of unit length, so that the components of $\vec{R}$
will be integers. We want to compute the mean value of $\vec{R}$ over
all polymers with $N$ steps
\begin{equation}
\langle R^{2}\rangle =\frac{\sum_{\vec{R_{m}^{N}}}z^{m}R^{2}}{\sum_{\vec{R_{m}^{N}}}
z^{m}},
\label{ea1}
\end{equation}
where $m$ is the number of bends in the walk and the sum
is over all configurations $\vec{R_{m}^{N}}$ of polymers with $N$
steps. Besides the first and last components the values of the other
$m-1$ components of $R$ are the numbers of steps between successive
bends in the walk. We should remember that  there are $q-2$
possibilities for each bend. So we may rewrite Eq. \ref{ea1}
\begin{equation}
\langle R^{2}\rangle =\frac{\sum_{m=0}^{N-1}a^{m}B_{N,m}}{\sum_{m=0}^{N-1}a^{m}A_{N,m}},
\label{ea2}
\end{equation}
where $a=(q-2)z$ embodies all dependence on coordination number and
statistical weight as long as $q\geq 4$, 
\begin{equation}
A_{N,m}=\sum_{\vec{R}_{m}^{N}}1,
\label{ea3}
\end{equation}
and
\begin{equation}
B_{N,m}=\sum_{\vec{R}_{m}^{N}}\sum_{i=0}^{m+1}R_{i}^{2}.
\label{ea4}
\end{equation}
Note that the effect of the bending energy can be described by introducing
an effective coordination number $q^\prime =a+2$ for an associated totally
flexible polymer.
The sums in $A_{N,m}$ and $B_{N,m}$ are over all possible values for
$\vec{R}$ with $m+1$ components and subjected to the constraint of the total number
of steps being equal to $N$, that is
\begin{equation}
\sum_{i=1}^{m+1}R_{i}=N.
\label{ea5}
\end{equation}
The sum in Eq. \ref{ea3} is just the number of vectors $\vec{R}$ with $m+1$
components which obey the constraint Eq. \ref{ea5}. Since the minimum value
of each component of $\vec{R}$ is equal to 1, it is convenient to
define $r_{i}=R_{i}-1$ and therefore $A_{N,m}$ is the number of ways
to put the $N-m-1$ remaining steps into the $m+1$ components of $\vec{R}$
\begin{equation}
A_{N,m}=\frac{(N-1)!}{m!(N-m-1)!}.
\label{ea6}
\end{equation}
The sum $B_{N,m}$ may then be rewritten as
\begin{equation}
B_{N,m}=\sum_{\vec{r}^N_m}\sum_{i=1}^{m+1}(1+r_{i})^{2},
\label{ea7}
\end{equation}
where each components $r_{i}$ assumes values between 0 and $N-m-1$
subject to the constraint of Eq. \ref{ea5}
\begin{equation}
\sum_{i=1}^{m+1}r_{i}=N-m-1.
\label{ea8}
\end{equation}
The calculation of $B_{N,m}$ is given in Appendix \ref{a2}, and the result
is
\begin{equation}
B_{N,m}=\frac{(m+1)(2N-m)N!}{(m+2)!(N-m-1)!}.
\label{ea9}
\end{equation}
Performing the sum in the denominator of Eq. \ref{ea2} taking Eq.
\ref{ea6} into account, we have
\begin{equation}
\langle R^{2}\rangle =\frac{N}{[1+a]^{N-1}}
\left[2(N+1)\sum_{m=0}^{N-1}\left(\begin{array}{c}
N-1\\m
\end{array}\right)\frac{a^{m}}{m+2}-\sum_{m=0}^{N-1}\left(\begin{array}{c}
N-1\\m
\end{array}\right)a^{m}\right].
\label{ea10}
\end{equation}
The first sum may be calculated by noting that
\begin{equation}
\int_{0}^{A}x(1+x)^{N-1}dx=A^{2}\sum_{m=0}^{N-1}\left(\begin{array}{c}
N-1\\m
\end{array}\right)\frac{A^{m}}{m+2},
\label{ea11}
\end{equation}
and therefore
\begin{equation}
\sum_{m=0}^{N-1}\left(\begin{array}{c}
N-1\\m
\end{array}\right)\frac{a^{m}}{m+2}=\frac{[1+a]^{N}[aN-1]+1}{N(N+1)a^{2}}.
\label{ea12}
\end{equation}
Substituting this result in Eq. \ref{ea10} and performing the second sum we
finally get the expression
\begin{equation}
\langle R^{2}\rangle
=\frac{2[1+a]}{a^{2}}\left[Na-1+\frac{1}{[1+a]^{N}}\right]-N.
\label{ea13}
\end{equation}

\section{DERIVATION OF $B_{N,m}$}
\label{a2}

In this appendix we want to derive Eq. \ref{ea9} for $B_{N,m}$.
Using Eq. \ref{ea8} and defining for convenience ${\cal N} =  N-m-1$ the
equation Eq. \ref{ea7} is rewritten as
\begin {equation}
B_{N,m}= (m+1) \sum_{j=0}^{{\cal N} +1}\frac{ ({\cal N} +m
-j)!j^2}{({\cal N} +1-j)! (m-1)!},
\label{eb1}
\end{equation}
Redefining the summation variable with
$i={\cal N} +1-j$ this equation turns out to be
\begin {equation}
B_{N,m}=(m+1)\sum_{i=0}^{\cal N} ( {\cal N}+1-i)^2\frac{(i+m-1)!}{i!(m-1)!}.
\label{eb2}
\end{equation}
Using the equality
\begin {equation}
\sum_{i=0}^{N}\frac{(m+i)!}{m!i!}=\frac{(m+N+1)!}{(m+1)!N!},
\label{eb3}
\end{equation}
we get after some manipulations
\begin {eqnarray}
B_{N,m}&=&(m+1)\left\{({\cal N}+1)\frac{({\cal N}+1+m)!}{{\cal N}!m!} 
-2({\cal N}+1)\frac{m({\cal N}+1+m)!}{{\cal N}!(m+1)!} \right.\nonumber\\
&&\left. \;\;\;\;\;\;\;\;\;\;\;\;\;\;\;\;+ \sum_{i=1}
^{{\cal N}+1} \frac{(i+m-1)!i}{(i-1)!(m-1)!}\right\}.
\label{eb4}
\end{eqnarray}

The last summation to be dealt with is just
\begin{equation}
\sum_{i=1}^{{\cal N} +1} \frac{(i+m-1)!i}{(i-1)!(m-1)!}.
\label{eb5}
\end{equation}
Defining $j=i-1$ it follows that
\begin{equation}
\sum_{j=0}^{{\cal N}} \frac{(j+m)!(j+1)}{j!(m-1)!} = \frac{m({\cal N}
+m +1)!}{{\cal N} !(m+1)!} + m(m+1)\frac{({\cal N}+m+1)!}{({\cal
N}-1)!(m+2)!}.
\label {eb6}
\end {equation}
Substitution in Eq. \ref{eb4} leads to
\begin{equation}
B_{N,m}= \frac {(m+1)({\cal N} +m +1)! [2{\cal N} +m +2]}{(m+2)!{\cal
N}!}.
\label{eb7}
\end{equation}

Substituting ${\cal N}=N-m-1$, we get
\begin{equation}
B_{N,m}=\frac{(m+1)(2N-m)N!}{(m+2)!(N-m-1)!}.
\label{eb8}
\end{equation}

\begin{figure}
\caption{A four-coordinated Cayley tree with a 2-step polymer
placed on it. The tree has $N_g=2$ generations, and is embedded 
on a cubic lattice. For the polymer shown $R^2=2$.}
\label{f1}
\end{figure}

\begin{figure}
\caption{The amplitude of $\langle R^2 \rangle$ as a function of $z$
and $y$ for a lattice with $s=1$ and $t=2$ ($q=6$). As expected, the
amplitude diverges as $z \rightarrow 0$. Since $s=1$, a divergence is
also observed as $y \rightarrow \infty$.}
\label{f2}
\end{figure}


\begin{references}

\bibitem[*]{l}On a leave from Departamento de F\'{\i}sica, Universidade
Federal de Santa Catarina.

\bibitem{f53}P.J.Flory, {\it Principles of Polymer Chemistry}(Cornell
University, Ithaca, New York,1953).

\bibitem{dg79}P.G. de Gennes, {\it Scaling Concepts in Polymer Physics}
(Cornell University Press, Ithaca, New York,1979).

\bibitem{n82}B. Nienhuis, Phys. Rev. Lett.{\bf 49},1062 (1982).

\bibitem{ts75}M.F. Thorpe and W.K.Scholl, J. Chem. Phys. {\bf 75}, 5143
(1981); W.Scholl and A.B. Thorpe, J. Chem. Phys. {\bf76}, 6386
(1982); J.W.Halley, H. Nakanishi, and R. Sundarajan, Phys.Rev.{\bf
B31}, 293 (1985); S. B. Lee and H. Nakanishi, Phys.Rev.{\bf B33},
1953 (1986); M.L. Glasser, V. Privman, and A. M. Szpilka, J. Phys.
{\bf A19}, L1185 (1986); V. Privman and S. Redner, Z.
Phys,{\bf B67},129 (1987); V. Privman and H.L.Frish, J. Chem. Phys.
{\bf 88}, 469 (1988); J.W.Halley, D. Atkatz, and H. Nakanishi, J.
Phys. {\bf A23}, 3297 (1990).

\bibitem{mn91}J. Moon and H. Nakanishi, Phys. Rev.{\bf A44}, 6427 (1991).

\bibitem{cfs92}C.J.Camacho, M.E.Fisher, and J.P.Straley, Phys.Rev.
{\bf A46}, 6300 (1992).

\bibitem{b82}R.J.Baxter, {\it Exactly Solved Models in Statistical
Mechanics} (Academic, london, 1982).

\bibitem{m92}F. Moraes, J. Physique {\bf I2}, 1657 (1992); F. Moraes,
Mod. Phys. Lett {\bf B8},909 (1994).

\bibitem{dq86}F. Peruggi, F. di Liberto, and G. Monroy, Physica
{\bf123A}, 175 (1984); S. L. A. de Queiroz, J. Phys. A {\bf19}, L433,
(1986).

\bibitem{hi98}C.-K. Hu and N. Sh. Izmailian, Phys. Rev. E {\bf 58},
1644 (1998).

\bibitem{sw87}J. F. Stilck and J. C. Wheeler, J. Stat. Phys. {\bf
46}, 1 (1987).

\bibitem{tm96}C. Tsallis and A. C. N. de Magalh\~aes, Phys. Rep. {\bf
268}, 305 (1996).
\end{references}
\end{document}